\documentclass[
    ,final            
  ]
  {aipproc}

\layoutstyle{6x9}

\usepackage{amssymb}
\usepackage{amsfonts}
\usepackage{amsmath}
\usepackage{epsfig}

\begin{document}

\title{Neutrino tri-bi-maximal mixing from $\Delta (27)$}


\author{I. de Medeiros Varzielas}{
  address={Rudolf Peierls Centre for Theoretical Physics,\\
University of Oxford, 1 Keble Road, Oxford, OX1 3NP}
}

\begin{abstract}
The observed neutrino mixing, having a near maximal atmospheric neutrino
mixing angle and a large solar mixing angle, is close to
tri-bi-maximal, putting leptonic mixing in contrast with the small mixing of the quark sector.
We discuss a model in which $\Delta(27)$ (a subgroup of $SU(3)$)
is the family symmetry, and tri-bi-maximal mixing directly follows from the vacuum
structure enforced by the discrete symmetry.
The model accounts for the observed quark and lepton masses
and the CKM matrix, as well as being consistent with an underlying
stage of Grand Unification.

Proceedings entry for SUSY 2006 \footnote{http://susy06.physics.uci.edu/talks/3/varzielas.pdf}, based on \cite{Ivo}.
\end{abstract}

\maketitle


\section{Introduction} \label{sec:Intro}

The observed neutrino oscillation parameters are consistent with a
tri-bi-maximal structure \cite{HPS}:%
\begin{equation}
U_{PMNS}\propto \left[ 
\begin{array}{ccc}
-\sqrt{\frac{2}{6}} & \sqrt{\frac{1}{3}} & 0 \\ 
\sqrt{\frac{1}{6}} & \sqrt{\frac{1}{3}} & \sqrt{\frac{1}{2}} \\ 
\sqrt{\frac{1}{6}} & \sqrt{\frac{1}{3}} & -\sqrt{\frac{1}{2}}%
\end{array}%
\right]  \label{eq:HPS}
\end{equation}

This simple form of leptonic mixing contrasts with the CKM
matrix. Family symmetries have been used to justify this structure, including some based on underlying $SU(3)$ family symmetry
\cite{SU(3)}.

$\Delta (27)$ is the semi-direct product group
$Z_{3} \ltimes Z_{3}'$ \cite{FFK}. This discrete group is interesting
as a family symmetry because it is the smallest subgroup of $SU(3)$ that has the useful
feature of having distinct triplets and anti-triplets (3 dimensional
irreducible representations).

\begin{table}[hb] \centering%
\begin{tabular}{|c|c|c|}
\hline
Field & $Z_{3}$ & $Z_{3}^{\prime }$ \\ \hline
$\phi_{1}$ & $\phi_{1}$ & $\phi_{2}$ \\ 
$\phi_{2}$ & $\alpha \phi_{2}$ & $\phi_{3}$ \\ 
$\phi_{3}$ & $\left( \alpha \right) ^{2}\phi_{3}$ & $\phi_{1}$ \\ 
\hline
\end{tabular}
\caption{} \centering
\label{ta:transformation}

\end{table}

The triplets $\phi_{i}$ transform as shown in TABLE
\ref{ta:transformation}. $\Delta (27)$ allows
all the $SU(3)$ invariants (being its subgroup) plus some additional invariants. Unlike smaller subgroups
like $\Delta (12)$ \cite{FFK}, it forbids invariants constructed solely from 2 triplets
(e.g. $\varphi \phi \equiv \varphi_{1} \phi_{1} + \varphi_{2} \phi_{2}
+ \varphi_{3} \phi_{3} $ is not invariant).

We can thus have left-handed fermions $\psi_{i}$ and their charge
conjugates $\psi^{c}_{i}$ share the same transformation properties under the family symmetry, which allows
straightforward embedding in GUTs (a desirable feature).

In fact, one can construct models having
the fermions (and their conjugates) transforming e.g. as triplets and the flavons responsable for breaking down
the family symmetry transforming as anti-triplets. The only allowed
terms that are quadratic in the fermions (triplets) arise when flavons
(anti-triplets) are included to make the appropriate family invariant
contractions, and these become mass terms for the fermions when the flavons acquire
non-vanishing vacuum expectation values (vevs).

\section{The model} \label{sec:Model}

The model aims to reproduce fermion masses and
mixings, with particular emphasis on reproducing neutrino
tri-bi-maximal mixing. We now discuss what is required in order to obtain
successful mass structures.

The model relies on the seesaw mechanism in order to obtain the
effective neutrino masses - so the heavy Majorana
neutrinos mass matrix plays a role in determining the effective
neutrino mass matrix (influencing the leptonic mixing).

A particularly interesting structure for the Majorana mass matrix is
one where there is strong hierarchy - Sequential Dominance (SD) scenarios (see \cite{SU(3)} and references therein).

Within SD, one can readily construct Yukawa structures that are phenomenologically viable for all
the fermions: the difference between the contrasting leptonic and
quark mixing is caused by the seesaw mediated intervention of the
strongly hierarchical Majorana masses, that can for example transform
the otherwise hierarchical dominance of the 3rd family
contribution characteristic of charged fermions, into a negligible
effect for the effective neutrino mass matrix.

The aim is to obtain viable quark structure as in
\cite{Roberts} and tri-bi-maximal mixing for the neutrinos after
seesaw has taken place. The charged leptons will introduce small corrections to
the neutrino mixing angles and yeld a near tri-bi-maximal PMNS matrix.
The most relevant Yukawa terms in the superpotential are as follows:

\begin{equation}
P_{Y}\sim\frac{1}{M^{2}}\bar{\phi}_{3}^{i}\psi_{i}\bar{\phi}%
_{3}^{j}\psi_{j}^{c}H  \label{eq:Y_P3_P3}
\end{equation}

\begin{equation}
+\frac{1}{M^{3}}\bar{\phi}_{23}^{i}\psi_{i}\bar{\phi}_{23}^{j}%
\psi_{j}^{c}HH_{45}  \label{eq:Y_P23_P23}
\end{equation}

\begin{equation}
+\frac{1}{M^{2}}\bar{\phi}_{23}^{i}\psi_{i}\bar{\phi}_{123}^{j}\psi_{j}^{c}H
\label{eq:Y_P23_P123}
\end{equation}

\begin{equation}
+\frac{1}{M^{2}}\bar{\phi}_{123}^{i}\psi_{i}\bar{\phi}_{23}^{j}\psi_{j}^{c}H
\label{eq:Y_P123_P23}
\end{equation}

$H$ stands for the usual SM or SUSY Higgs, and $H_{45}$ is an additional scalar
field whose role is discussed in \cite{Ivo}.

The flavon vevs that yeld the desired masses are:

\begin{equation}
\left\langle \bar{\phi}_{3}\right\rangle=\left(%
\begin{array}{ccc}
0,0,a
\end{array} \right)  \label{eq:P3 vev}
\end{equation}

\begin{equation}
\left\langle \bar{\phi}_{23}\right\rangle=\left(%
\begin{array}{ccc}
0,-b,b
\end{array}%
\right)  \label{eq:P23 vev}
\end{equation}

\begin{equation}
\left\langle \bar{\phi}%
_{123}\right\rangle=\left( 
\begin{array}{ccc}
c,c,c
\end{array}%
\right)  \label{eq:P123 vev}
\end{equation}

The combination of the $P_{Y}$ terms with these $\bar{\phi}$ vevs leads to
similar textures to those in \cite{Roberts} for quarks, and a charged
lepton mass matrix very similar to that of the down quarks (i.e. with small
mixing).

It is important to discuss how the vevs of eqs.(\ref{eq:P3 vev}) to (\ref{eq:P123 vev}) arise. One way of aligning the vevs relies solely on soft
terms. The key there is that the symmetry is discrete and thus breaks
the continuum of vacuum states. A simple example of how that can arise
is through soft quartic terms proportional to $m^{2}_{3/2}
\bar{\phi}^{\dagger i} \bar{\phi}_{i}
\bar{\phi}^{\dagger i} \bar{\phi}_{i}$. This type of term is not allowed by $SU(3)$,
but is allowed by $\Delta(27)$. If $\bar{\phi}$ acquires a non-zero vev, the
minimisation of the quartic term determines the direction of the vev to be that of
eq.(\ref{eq:P3 vev}) - when the coefficient of the quartic term is negative;
or eq.(\ref{eq:P123 vev}) - when the coefficient of the quartic term is
positive. Relative alignment as in eq.(\ref{eq:P23 vev}) comes from
mixed terms involving e.g. $\bar{\phi}_{23}^{i} \bar{\phi}_{123 i}^{\dagger}$. The complete alignment is discussed in detail in \cite{Ivo}.

The $\Delta (27)$ family symmetry isn't enough in order to ensure that
only the desired $P_{Y}$ terms are allowed. We introduce additional symmetries to further constrain terms that can spoil the mass
structures.

\begin{table}[hb] \centering

\begin{tabular}{|c|c|c|c|}
\hline
Field & $\Delta (27)$ & $U(1)$ & $R$ symmetry \\ \hline
$\psi $ & $\mathbf{3}$ & $\mathbf{0}$ & $\mathbf{1}$ \\ 
$\psi ^{c}$ & $\mathbf{3}$ & $\mathbf{0}$ & $\mathbf{1}$ \\ 
$H$ & $\mathbf{1}$ & $\mathbf{0}$ & $\mathbf{0}$ \\ 
$H_{45}$ & $\mathbf{1}$ & $\mathbf{2}$ & $\mathbf{0}$ \\ 
$\bar{\phi}_{3}$ & $\bar{\mathbf{3}}$ & $\mathbf{0}$ & $\mathbf{0}$ \\ 
$\bar{\phi}_{23}$ & $\bar{\mathbf{3}}$ & $\mathbf{-1}$ & $\mathbf{0}$ \\ 
$\bar{\phi}_{123}$ & $\bar{\mathbf{3}}$ & $\mathbf{1}$ & $\mathbf{0}$ \\ \hline
\end{tabular}

\caption{} \centering

\label{ta:charges}

\end{table}

The charge assignments in TABLE \ref{ta:charges} give rise to the superpotential in
eqs.(\ref{eq:Y_P3_P3}) to (\ref{eq:Y_P123_P23}), forbidding unwanted terms.

The symmetries may be extended to the sector giving rise to Majorana masses as described in \cite{Ivo}.
They fulfill SD and combine with the
specific structure of the Yukawa terms (namely
eq.(\ref{eq:Y_P23_P123}) and eq.(\ref{eq:Y_P123_P23})) to yeld a
tri-bi-maximally mixed effective neutrino
mass matrix.

\section{Conclusion} \label{sec:Conclusions}

After obtaining the tri-bi-maximal neutrino mixing, it's necessary to take
into account the charged lepton mixing in order to get the PMNS
angles \cite{Antusch}. The charged lepton mass matrix was also
obtained from the model, and so we conclude that the model predicts
the following leptonic mixing angles \cite{Ivo}:

\begin{equation}
\sin^{2}\theta_{12} = \frac{1}{3}\pm_{0.048}^{0.052}
\end{equation}

\begin{equation}
\sin^{2}\theta_{23} = \frac{1}{2}\pm_{0.058}^{0.061}
\end{equation}

\begin{equation}
\sin^{2}\theta_{13} = 0.0028
\end{equation}

In conclusion, the model is phenomenologically viable, and is
consistent with underlying Grand Unification.

The model relies on the seesaw mechanism under a SD scenario, and also on
misalignment of vevs (in this case through relatively simple soft
terms allowed by the family symmetry). The tri-bi-maximal mixing is, in that sense, directly related
to the discrete group $\Delta (27)$.


\begin{theacknowledgments}
\noindent I. de Medeiros Varzielas was supported by FCT under the grant
SFRH/BD/12218/2003.

\noindent I. de Medeiros Varzielas is grateful to Graham Ross for
helpful comments and discussion.
\end{theacknowledgments}



\bibliographystyle{aipproc}   

\end{document}